\documentclass[sigconf]{acmart}
\usepackage{balance}
\def\BibTeX{{\rm B\kern-.05em{\sc i\kern-.025em b}\kern-.08emT\kern-.1667em\lower.7ex\hbox{E}\kern-.125emX}}
    
\copyrightyear{2019}
\acmYear{2019}
\acmConference[MM '19]{Proceedings of the 27th ACM International Conference on
Multimedia}{October 21--25, 2019}{Nice, France}
\acmBooktitle{Proceedings of the 27th ACM International Conference on Multimedia (MM '19),
October 21--25, 2019, Nice, France}
\acmPrice{15.00}
\acmDOI{10.1145/3343031.3350918}
\acmISBN{978-1-4503-6889-6/19/10}

\begin{document}

\fancyhead{}
  % do not delete this code.

% The "title" command has an optional parameter, allowing the author to define a "short title" to be used in page headers.
\title{Informative Visual Storytelling with Cross-modal Rules}

% The "author" command and its associated commands are used to define the authors and their affiliations.
% Of note is the shared affiliation of the first two authors, and the "authornote" and "authornotemark" commands
% used to denote shared contribution to the research.


%\email{lijiacheng@zju.edu.cn }

\author{Jiacheng Li}
\affiliation{%
  \institution{College of Computer Science and
Technology, ZheJiang Universety}
  %\city{HangZhou}
  %\country{China}
}
\email{lijiacheng@zju.edu.cn}

\author{Haizhou Shi}
\affiliation{%
    \institution{College of Computer Science and
Technology, ZheJiang Universety}
  %\city{HangZhou}
  %\country{China}
}
\email{shihaizhou@zju.edu.cn}

\author{Siliang Tang}
\authornote{Siliang Tang is the corresponding author.}
\affiliation{%
    \institution{College of Computer Science and
Technology, ZheJiang Universety}
  %\city{HangZhou}
  %\country{China}
}
\email{siliang@zju.edu.cn}

\author{Fei Wu}
\affiliation{%
  \  \institution{College of Computer Science and
Technology, ZheJiang Universety}
  %\city{HangZhou}
  %\country{China}
}
\email{wufei@cs.zju.edu.cn}

\author{Yueting Zhuang}
\affiliation{%
    \institution{College of Computer Science and
Technology, ZheJiang Universety}
  %\city{HangZhou}
  %\country{China}
}
\email{yzhuang@cs.zju.edu.cn}

%
% By default, the full list of authors will be used in the page headers. Often, this list is too long, and will overlap
% other information printed in the page headers. This command allows the author to define a more concise list
% of authors' names for this purpose.

%
% The abstract is a short summary of the work to be presented in the article.
\begin{abstract}
Existing methods in the Visual Storytelling field often suffer from the problem of generating general descriptions, while the image contains a lot of meaningful contents remaining unnoticed. The failure of informative story generation can be concluded to the model's incompetence of capturing enough meaningful concepts. The categories of these concepts include entities, attributes, actions, and events, which are in some cases crucial to grounded storytelling. To solve this problem, we propose a method to mine the cross-modal rules to help the model infer these informative concepts given certain visual input. We first build the multimodal transactions by concatenating the CNN activations and the word indices. Then we use the association rule mining algorithm to mine the cross-modal rules, which will be used for the concept inference. With the help of the cross-modal rules, the generated stories are more grounded and informative. Besides, our proposed method holds the advantages of interpretation, expandability, and transferability, indicating potential for wider application. Finally, we leverage these concepts in our encoder-decoder framework with the attention mechanism. We conduct several experiments on the VIsual StoryTelling~(VIST) dataset, the results of which demonstrate the effectiveness of our approach in terms of both automatic metrics and human evaluation. Additional experiments are also conducted showing that our mined cross-modal rules as additional knowledge helps the model gain better performance when trained on a small dataset.
\end{abstract}




%
% Keywords. The author(s) should pick words that accurately describe the work being
% presented. Separate the keywords with commas.

%
% This command processes the author and affiliation and title information and builds
% the first part of the formatted document.
\maketitle

\section{Introduction}

With the tremendous success achieved in image captioning recently, the researchers have taken a step further into the task of visual storytelling, which requires the model to generate a coherent and narrative story given a photo stream. The task of visual storytelling is challenging mainly due to the following reasons. Firstly, compared to image captioning, the large visual variance of photo streams demands a more exquisite visual modeling technique. Secondly, stories could be subjective as they may contain rich emotions, imaginations, latent topics and events, and distinctive narrative threads which are largely decided by annotators' personal experiences. This even requiring the model to be capable of inferring concepts that do not explicitly appear in images.

\begin{figure}[htbp]
\centering 
\includegraphics[height=3.69cm, width=8.5cm]{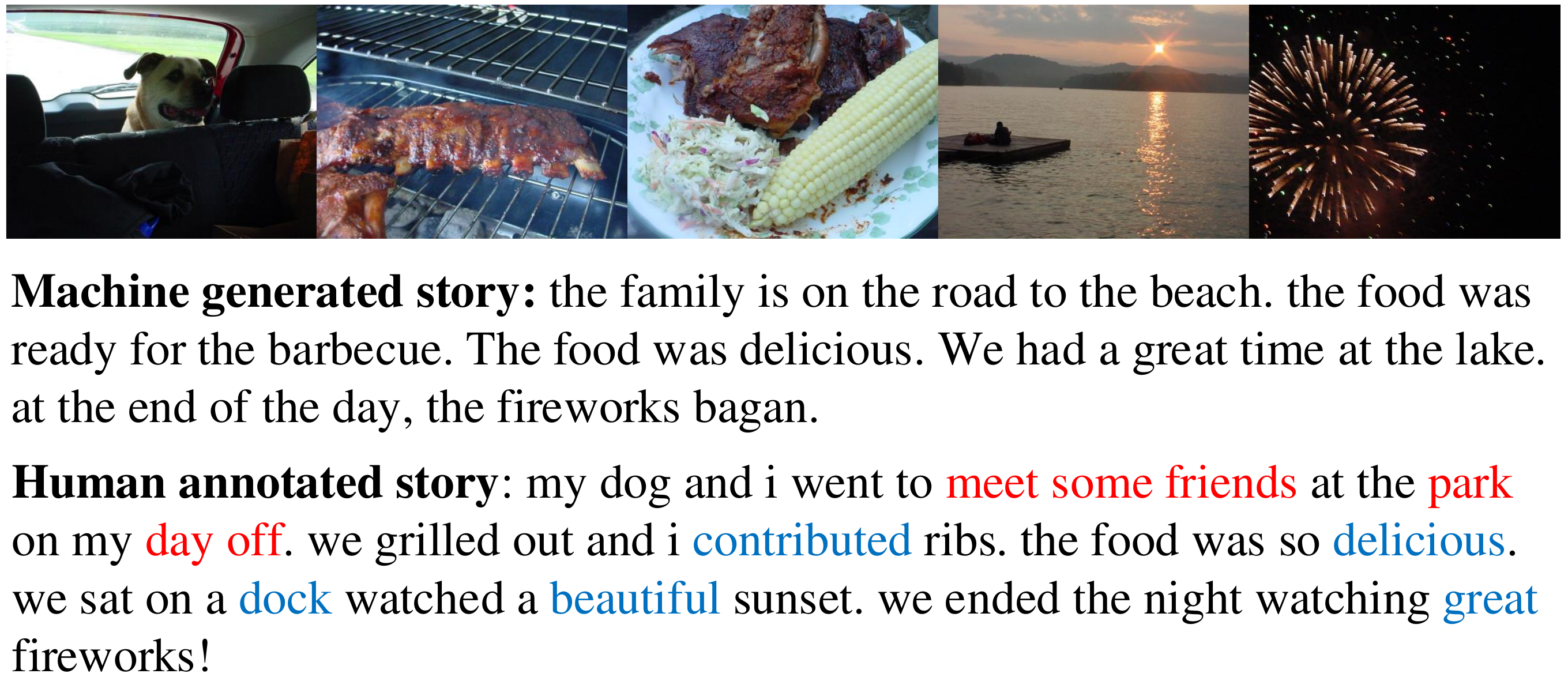} 
\caption{An example of a machine generated story and a human annotated story. The words in red are imagination which is out of vision but related; the words in blue are not explicitly shown but could be inferred from vision.} 
\label{figure1}
\end{figure}

Previous studies~\cite{huang2016visual, Kim18, WangFu18, Huang18, wang2018no, Hsu18} are mainly inspired by the success of sequence-to-sequence translation structure, which translates the image sequence into a sequence of sentences. However, these approaches depend merely on the global visual features extracted by networks such as Convolutional Neural Network~(CNN), so that they tend to give general and uninformative narratives. Although~\cite{WangHanQi17} has realized the problem and attempt to fix the problem by attending to various salient regions of images to form local features, it still fails to infer intricate concepts such as actions, topics, and events which may be crucial to storytelling, as shown in Figure~\ref{figure1}.

In order to solve the aforementioned problem, we propose a method called VSCMR~(Visual Storytelling with Cross-Modal Rules) to build cross-modal rules and leverage them in the current sequence-to-sequence framework. Inspired by the mid-level pattern mining algorithm~\cite{li2015mid}, we first create cross-modal transactions~(also known as records in some other context) by combining CNN filter responses of images and the indices of filtered words in corresponding annotations of images. Then, we adopt the association rule mining algorithm to discover the rules which can be viewed as a mapping from visual patterns to semantic concepts~\cite{agrawal1993mining}. These rules reveal the relationships between the CNN activations and the concepts based on their co-occurrence frequency. The mined rules are easy to expand and transfer: whenever there is a new set of pairwise image-description data, we could simply mine the cross-modal rules on the new dataset and add them into the existing rules to form a larger set of rules.

During story generation, we directly apply the mined rules to the filters' responses of images to get semantic concepts, which in our case, are a set of words. Inspired by~\cite{you2016image}, we use the semantic attention mechanism to embed the mined rules into our sequence-to-sequence framework. We implement the semantic attention score by calculating the correlation between the hidden states of the Recurrent Neural Network~(RNN) and the word embedding of all semantic concepts in the photo stream. The attended context vector is the weighted average of semantic concepts. The context vector, which serves as a heuristic cue, will be concatenated to the input of the RNN. By using cross-modal rules and attention mechanism, our proposed framework could be transferred to any other end-to-end model that focuses on other cross-modal applications.

In summary, our approach has three major advantages. 
    First of all, our proposed cross-modal rules are capable of inferring a wide range of concepts including concrete entities, attributes, and actions. It can also speculate the topic or event out of given visual input since the annotations may contain these highly abstract notions. What's more, these inferred concepts are highly correlated with specific visual patterns, such as "delicious" and "food", "beautiful" and "sunset"~(as shown in Figure \ref{figure1}), which is conducive to a more informative and interpretable storytelling manner.
        Secondly, the rules are portable and expandable. The effectiveness of this method relies on the fact that once the rules are established, they can be directly joined into another set of rules without reconstruction, which saves tons of resources and time. 
            Last but not the least, by making full use of existing storytelling annotations, our cross-modal rules are mined in a weakly supervised manner, which does not need extra human labels and could be easily transferred to any cross-modal dataset as long as it has a weak mapping between data of two modals.

We carry out experiments on the VIST dataset~\cite{huang2016visual} and our proposed method of cross-modal rules shows performance gains on all of the automatic metrics. Being aware of the limitation of automatic metrics, we also estimate our model by human evaluation, the result of which shows that the stories generated with cross-modal rules are more relative, expressive and concrete. Further experiments also prove that our mined cross-modal rules as additional transferable knowledge always helps the model gain better performance especially when there are only relatively few training examples are available. 

Our main contributions are as follows:
\begin{itemize}
    \item We propose to use cross-modal rules in the visual storytelling task. Our proposed VSCMR has the capability of capturing visual grounded concepts including entities, attributes, actions, topics, and events, leading to a more informative storytelling manner. It demonstrates the capabilities of better interpretation, expandability, and transferability. 
    \item We incorporate high-level concepts into the storytelling model as heuristic cues with semantic attention for better generation. 
    \item We conduct experiments on VIST dataset and evaluate the model by both automatic metrics and human evaluation, the results of which demonstrate the effectiveness of our approach.
\end{itemize}

\section{RELATED WORK}
\subsection{Visual Storytelling}
The visual storytelling task is usually solved by the sequence-to-sequence learning framework. For example, \cite{huang2016visual} encodes the image stream into one fixed-length vector and decodes the vector into a sequence of words; \cite{WangHanQi17} propose a Contextual Attention Network(CAN) which encodes the image stream by attending to various salient regions and contextualize them according to their mutual similarity, then it decodes these representations into story using the hierarchical GRU; \cite{Hsu18} and \cite{Kim18} adopt a similar base architecture: bidirectional LSTM encodes the visual content and feeds the visual representation into the LSTM decoder at each time step, while~\cite{Hsu18} uses inter-sentence diverse beam search to explicitly reduce redundancy during the story generation. 

Researchers also explore the possibilities of applying more advanced techniques such as RL and GAN to enhance the sequence-to-sequence framework. \cite{WangFu18} propose a reinforcement learning framework with a cross-modal reward and a language-style reward, which enables hierarchical training policy for the story-style paragraph generation. \cite{wang2018no} develop and integrate the policy and the reward models to adversarially train the agent thus it can freely explore the strategy of generating the text. 

In summary, most of the methods in visual storytelling focus more on the word decoding policy learning than the grounded visual perception, while our work leverages the fine-grained visual information and explicitly infer the semantic concepts for given visual input, enabling a more informative storytelling style. 

\subsection{Association Rule Mining}
The association rule mining was first proposed in the commercial scenarios where the company wants to understand the customers' behavior.By observing and mining the regularities between products,it is possible to improve the pricing policy and product placement~\cite{agrawal1993mining}. In the context of pattern mining, an association rule between an itemset $I$ and $J$ implies that a transaction containing $I$ is also likely to contain $J$. For example, the association rule between the itemset $\{\text{"milk"}, \text{"diaper"}\}$ and $\{\text{"beer"}\}$ is denoted as $\{\text{"milk"}, \text{"diaper"}\} \Rightarrow \{\text{""beer"}\}$, indicating that the customer who buys the milk and the diaper is likely to buy the beer as well.

This approach is adopted to find the underlying association between items in many fields: \cite{Esfahani16} proposes a tree-based association rule mining algorithm to detect anomalies in Web services. \cite{Kong16} developed a rare association rule mining algorithm for the network intrusion. In the field of computer vision, \cite{li2015mid} use this technique to mine and to cluster the visual patterns, which prove to be conducive to the downstream tasks.  

\subsection{Incorporating Attributes using Attention}
The attention mechanism was first proposed in neural machine translation to help the model focus on the apposite positions of the source text. It provides the RNN model with dynamic contextual features based on the previous result~\cite{Bahdanau14, luong2015effective, Firat16}. After that, the mechanism is adopted in the domain of vision-to-text, which helps the agent to selectively focus on the relevant regions during generation: by attending to the extracted features of the visual patches, the generation of the model is more fact-based and fine-grained~\cite{Bahdanau14, xu2015show, Chen17}. 

The mechanism and its variants are instinctive and efficacious, while they fail to explain the interrelation between the visual content and the abstract concept such as "delicous" and "beautiful". To fix this defect, a series of methods are proposed to avoid direct visual-to-text translation. Instead, they use the CNN network only to extract the relevant words (also named as visual attributes) from the visual input and exploit them in the downstream generation as the pure source or ancillary information~\cite{chen2018show, yao2017boosting, you2016image}. \cite{fang2015captions} use multi-instance learning to detect the words and use the maximum-entropy model to generate the caption. \cite{yao2017boosting} use the same method to detect words while they only leverage the final probability distribution over the vocabulary as the attributes. \cite{you2016image} propose the semantic attention mechanism to dynamically attend to both concrete objects and abstract concepts: they train the word detectors by the weakly supervised learning method and refer to the detected words at each time step as extra information. 

\subsection{Comparison with MDPM}
Our work is inspired by \cite{li2015mid} but is different from \cite{li2015mid}. They only add 'pos' and 'neg' tag into the transaction for pattern recognition while we incorporate the whole description to build the multimodal transaction. In this way, we extend the bi-classification approach to a many-to-many mapping which can be used for concepts inference.  Besides, considering that the potential search space grows exponentially with the transaction, we further design a part of speech filter and a lemmatization filter to shorten the transaction. We also replace the apriori alogorithm with the more efficient FP-Growth algorithm which builds a FP-tree by inserting transactions into a trie.

\begin{figure*}[t] 
\centering 
\includegraphics[height=5.78cm, width=17.78cm]{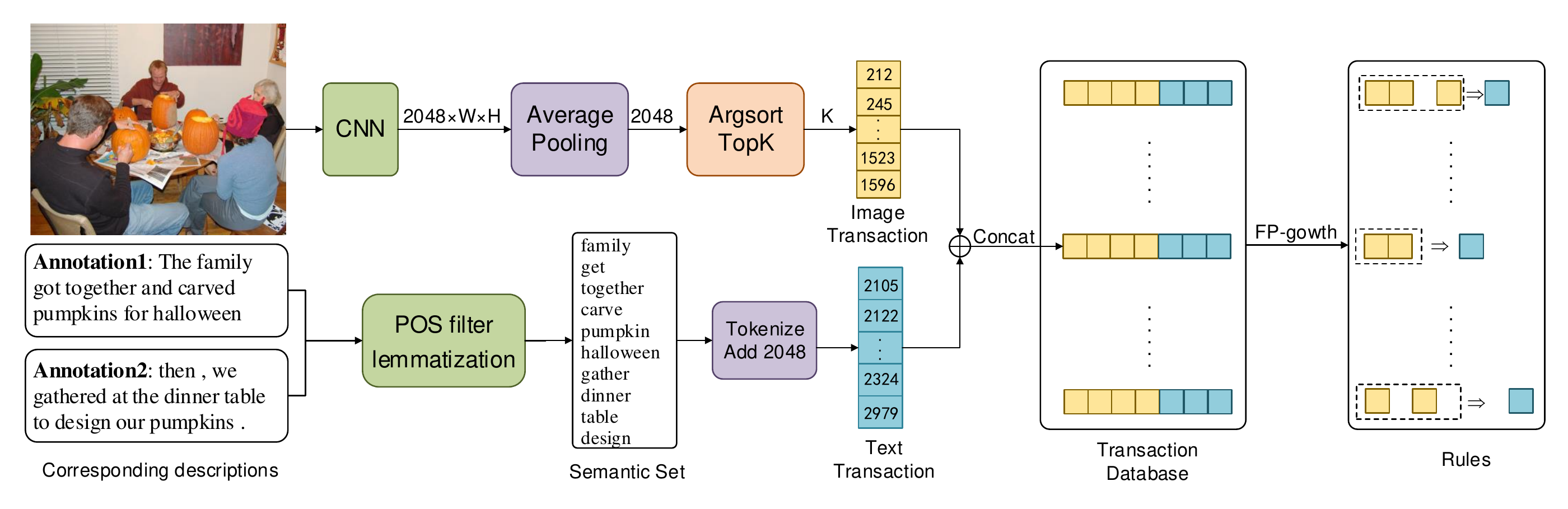} 
\caption{An overview of cross-modal rule mining. We extract image features by CNN without resizing at input. An average pooling layer is followed to get the activation for each filter. The image transaction is represented by the dimension indices of the k largest magnitudes of CNN activation. This is implemented by keeping the top K element in the argsort result of the CNN activation. For all the corresponding description of the image, we filter out non-semantic words according to the part-of-speech and implement lemmatization on each word. We represent the text transaction by the indexes of words plus 2048. Then the image transaction and text transaction are concatenated to a cross-modal transaction. Finally, we apply the Fp-Growth\cite{Han:2000:MFP:335191.335372} algorithm to the transaction database to find cross-modal rules for semantic concept inference. }
\label{figure2}
\end{figure*}

\section{APPROACH}
\subsection{Semantic Concept Inference Framework\label{section3.1}}
In this section, we give out details of concept inference method. 
\subsubsection{Cross-modal Transaction Creation}
 Transaction is a basic concept in association rule mining. Let $A=\{a_1, a_2, ..., a_n\}$ be a set of $n$ binary attributes called items. Let $D=\{t_1, t_2, ..., t_m\}$ be a set of transactions called the database. Each transaction in $D$ contains a subset of the items in $A$. Each item is recorded by a binary value: $1$ represents the presence of the item in the corresponding transaction while $0$ represents the absence. The transactions are represented by the indices of present items for convenience.
 For computability, a transaction should satisfy the following requirements:
\begin{itemize}
    \item The number of items in each transaction should be small as the potential search space grows exponentially with it.
    \item The record in each transaction must be a set of integers as opposed to the real-value elements of most vision features.
\end{itemize}

The key to our method is the creation of the cross-modal transactions. Our rule mining pipeline is illustrated in Figure \ref{figure2}. In order to get cross-modal rules for inference, each transaction is composed of the image transaction and the text transaction. 

We begin by extracting features of images with Resnet152\cite{he2016deep}. We don't resize image during input and the output shape of conv5\_x from Resnet152 is $W\times H \times 2048$, where $W$ and $H$ is dependent on the input size of the image. Then we imply 2D average pooling to get a 2048-dimension vector for each image and each dimension of the vector represents the activation of each filter. Following the work of \cite{li2015mid}, we treat each dimension index of CNN activation as an item(2048 in total). In their work, they find appealing properties of CNN activation and conclude that the discriminative information with a image's CNN activation is mostly embedded in the dimension indices of the k largest magnitudes. So, in order to satisfy the two requirements described before, we only keep top-k dimension indices as image transaction for each image.

For the text transaction part, we first put words of all corresponding descriptions of a single image into a set to avoid word repetition. We only keep semantic words such as verb, adjective, adverb by part of speech filtering. Then we implement lemmatization on these words to eliminate the effects of verb tenses, the plural form of nouns and comparative degree of adjectives and adverbs. Each word in the set is treated as an item. We represent the text transaction by the indices of these words in the vocabulary.

Finally, the image transaction and the corresponding text transaction are concatenated to form a cross-modal transaction. Each number in the text transaction are added 2048 to avoid overlapping of the indexes. For example, given a CNN activation of an image which has 3 largest magnitudes in its 18th, 1996th dimension and filtered words indexes 18, and 100, the corresponding transaction  will be \{18, 1996, 2066, 2148\}.

\subsubsection{Cross-Modal Rule mining} For better description, we first introduce the basic concepts in association rule mining. 

\textbf{Association rules}
 An association rule is defined as an implication of the form: 
 \begin{equation}
      X\Rightarrow a_j, X\subseteq{A}, a_j \in{A}
 \end{equation}
 where $X$ is a subset of $A$ and $a_j$ is a single item. The rule means that if itemset $X$ appears in the dataset, item $a_j$ is also likely to appear.
 
In order to select interesting rules from the set of all possible rules, we set minimum thresholds on support and confidence.

\textbf{Support} Support is an indication of how frequently the itemset appears in the dataset. The support of $X$ with respect to $D$ is defined as the proportion of transaction $t$ in the dataset  which contains the itemset $X$:
\begin{equation}
    \text{supp}(X)=\frac{|t\in D;x\subseteq{t}|}{|D|}
\end{equation}
where |$\cdot$| measures the cardinality of a set. $I$ is called a frequent itemset when supp($I$) is larger than a predefined threshold.

\textbf{Confidence} Confidence is an indication of how often the rule has been found to be true. The confidence value of a rule is the proportion of the transactions that contains  $X$ which also contains $a_j$:
\begin{equation}
   \text{conf}(X\Rightarrow a_j)=\frac{\text{supp}(X\cup a_j)}{\text{supp}(X)}        
\end{equation}
where the fraction between the cardinality of transactions is equivalent to the fraction between support of the transactions 
 
\textbf{Association Rule Mining}
After building the transaction database, we use Fp-Growth algorithm\cite{han2000mining} to discover association rules satisfying the following two criteria: 
\begin{align}
    \text{supp}(X) &>\text{supp}_{min} \\
    \text{conf}(X\Rightarrow a_j) &> \text{conf}_{min}
\end{align}
where $\text{supp}_{min}$ and $\text{conf}_{min}$ are the predefined thresholds for support and confidence. We obtain rules mapping from vision to text by further constraints on rules $X\Rightarrow a_j$:
\begin{align}
    a_k &< 2048, \forall a_k \in X \\
    a_j &\geq 2048
\end{align}
where the inequation~(6)(7) constrain that all the items in $X$ come from image transaction and the inferred item is a concept word.
In this way, the rule $X\Rightarrow a_j$ denotes that if all the filters in set $X$ are activated, word $a_j$ is inferred. Given an image, we infer semantic concepts by checking its image transaction with all mined rules. 

\begin{figure*}[t] 
\centering 
\includegraphics[height=7.05cm, width=15.248cm]{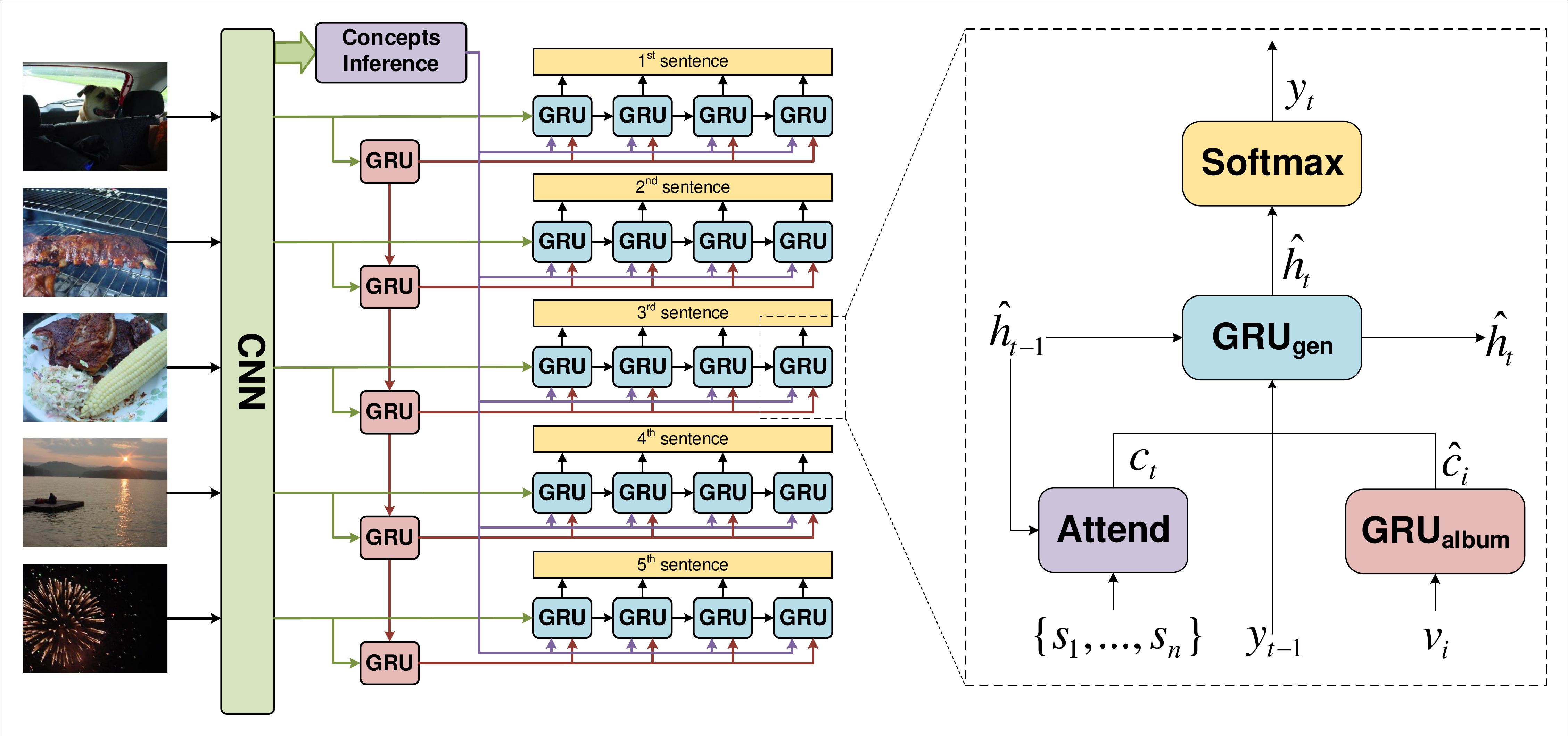} 
\caption{An overview of our story generation model. It contains an album GRU in red and a generation GRU in blue. We first extract image features with the Resnet152 and these features are used for the concepts inference and the generation GRU initialization. The features are also encoded to the visual context vectors and the inferred semantic concepts are fused to the semantic context vector. The input of the generation GRU is the concatenation of the last word, the visual context vector, and the semantic context vector. Finally, we concatenate all the generated sentences to form a story.} 
\label{figure3}
\end{figure*}

\subsection{Story Generation Model}
\subsubsection{Overall Framework}
Figure \ref{figure3} shows an overview of our generation model. The main part of our story generation model is a decoder composed of the hierachical GRUs with attention mechanism. Given a photo stream of $I=\{I_1,I_2,I_3,I_4,I_5\}$, we capture image features $V=\{v_1,v_2,v_3,v_4,v_5\}$ with the Resnet152. The features are the 2048-dimension output vectors of average pooling on conv5\_x layer. It is worth noting that the image features $V$ will be reused many times.

We first utilize a BiGRU called album GRU to encode the high-level image features $V$ to generate global visual context vectors $\hat{c}$. The visual context vectors capture the relations among images by encoding the image features. Then we utilize a GRU called generation GRU to generate word sequences. The hidden state $\hat{h}_{i,0}$ of the generation GRU is initialized by the corresponding visual context vector:
\begin{align}
    h_i &= \text{BiGRU}_{album}(V) = [\overleftarrow{h_i};\overrightarrow{h_i}]\\
    \hat{c}_i &= \text{Relu}(W_v h_i+b_v).\\
    \hat{h}_{i,0} &= W_0\hat{c}_i+b_0
\end{align}
where $h_i\in\mathbb{R}^{d\times 1}$ is the hidden state of the album GRU, d is the hidden size, [;] denotes concatenating, $\hat{h}_{i,0}$ is the initial hidden state of the generation GRU, $w_v\in\mathbb{R}^{d\times d}$ and $b_v\in\mathbb{R}^{d\times 1}$ are learned parameters, each hidden state $\hat{h}_{i,0}$ will be used to generate a sentence $Y_i (i=1,...,5)$, then the generated sentences are concatenated to form a story. The detailed decoding step will be described later. 

\subsubsection{Decoding Step}
The image features $V$ are also used for the image transaction creation and semantic concept inference as described in section \ref{section3.1}. Each semantic concept is represented by a one-hot vector $s$. We put all the inferred semantic concepts of a story into a set $S=\{s_1,..., s_n\}$ to eliminate redundancy. Then we get the semantic context vector $c_t$ by the following formula:
\begin{align}
    a_{t,j} &= \text{softmax}({\hat{h}_{i,t-1}}^TU(Es_j)) \\
    c_t^* &= \sum_{k=1}^{n}a_{t,j}(Es_j) \\
    c_t &= \text{tanh}(W_cc_t^*+b_c)
\end{align}
where $E\in\mathbb{R}^{e\times v}$ is the word embedding matrix, here $e$ is the embedding size and $v$ is the vocabulary size, $\hat{h}_{i,t-1}\in\mathbb{R}^{h\times 1}$ is the hidden state of sentence generation GRU, the attention score $a_{t,i}$ for semantic concept $s_i$ is calculated by a bilinear matrix $U\in\mathbb{R}^{h\times e}$. The final semantic context vector is activated by a tanh function. The matrices E, U, $W_c$ and the bias $b_c$ are all learned parameters.

At decoding time, the $i$th hidden state will be decoded to the ith sentence. The hidden state of the generation GRU is updated by:
\begin{align}
    \hat{h}_t &= \text{GRU}_{gen}(\hat{h}_{t-1}, x_t) \\
    x_t &= [y_{t-1};\hat{c};c_t] \\
    p(y_t) &= \text{softmax}(\text{tanh}(W_h\hat{h}_t+b_h))
\end{align}
where we leave out the subscript of $i$ for the simplicity of the formula, $y_{t-1}$ is the word embedding of the last word. The $\hat{c}$ and $c_t$ are the visual context vector and the semantic context vector, we concatenate the 3 vectors together as the input for the generation GRU. $\hat{h}_0$ contains the top-down information, $c_t$ contains the bottom-up information and the $\hat{c}$ captures the relations among images. In this way, the model is free to choose which kind of information to be used during generation. $p(y_t)$ is the probability distribution of the next word, $W_h$ projects the hidden state to the vocabulary size, the next word is selected by beam search.

\section{EXPERIMENTS}
\subsection{Experimental Setup}

\textbf{Dataset} The VIST dataset~\cite{huang2016visual} is the first dataset for the sequential vision-to-language task including visual storytelling, which contains 10,117 Flickr albums with 210,819 unique images. Each story consists of 5 images selected from an album and 5 corresponding description as a story. The images are selected, arranged and annotated by AMT workers and the same album is paired with 5 different stories as reference. After filtering out broken images, there are 50,136 samples split into 3 parts: 40,098 for training, 4,988 for validation and 5,050 for testing.

\textbf{Training} We extract image features with pre-trained Resnet152 model~\cite{he2016deep} by applying average pooling on the output of conv5$\_$x layer.  We represent words occurring fewer than 3 times as one special word UNK and build a vocabulary of size 9,837. Only 10 largest magnitudes in CNN activation are kept as image transaction since the number of the items in a transaction should be small. The thresholds for support and confidence will be discussed in next subsection. Note that only training set is used for rule mining. During story generation, the hidden state dimension of the GRUs is set to 512. The 512-dimension word embedding are learned during training. At decoding time, beam search with beam size of 3 and schedule sampling\cite{bengio2015scheduled} are adopted.

\begin{figure*}[t] 
\centering 
\centerline{\includegraphics[height=5.72cm, width=16.3cm]{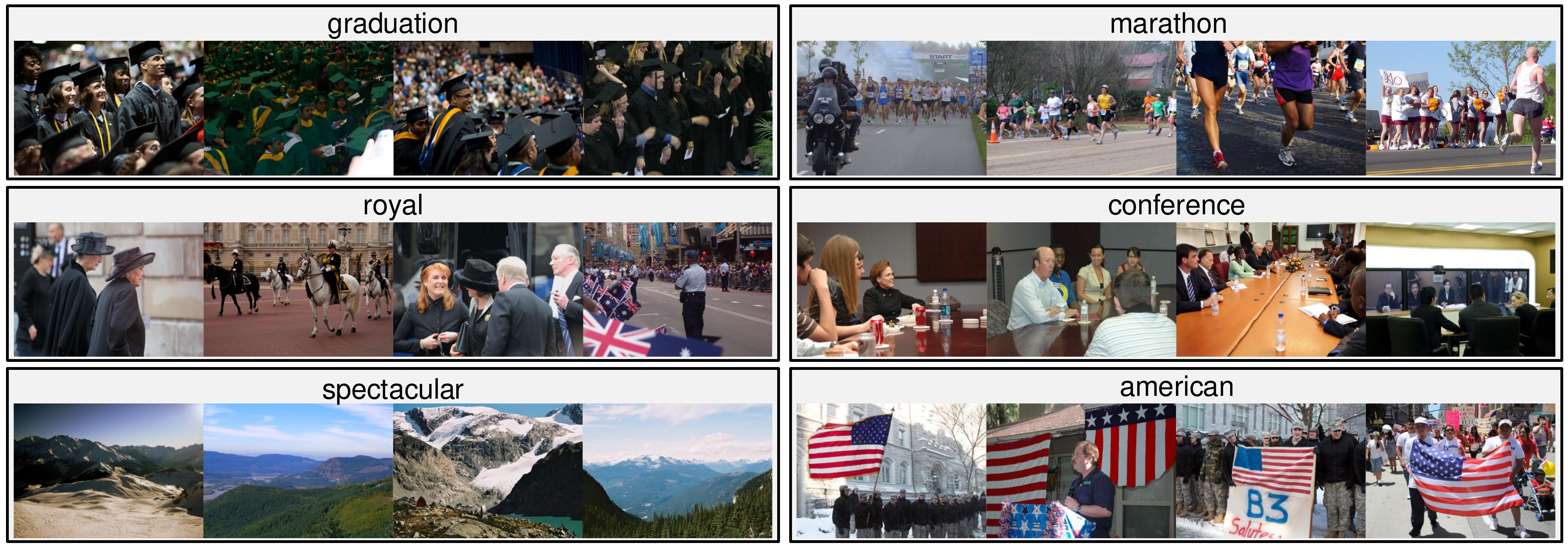}} 
\caption{The abstract concepts inferred from the cross-modal rules.} 
\label{figure4}
\end{figure*}

\begin{figure*}[t] 
\centering 
\centerline{\includegraphics[height=4.74cm, width=16.3cm]{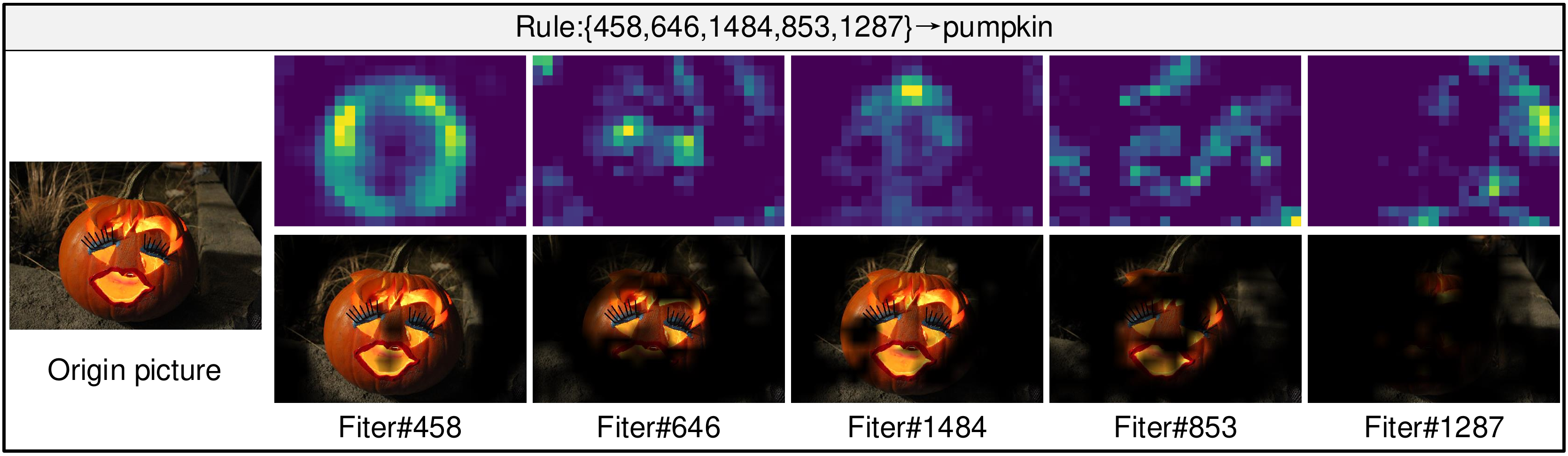}} 
\caption{Visualization of activated filters in the rule. The upper row is the visualization of feature maps of the corresponding filters. The lower row is generated by applying the interpolation of the feature map as a mask to the original picture.} 
\label{figure5}
\end{figure*}

\begin{table}[t]
\centering
\caption{Evaluation on rules with different thresholds. $\uparrow$ indicates larger better; $\downarrow$ indicates smaller better.}
\begin{tabular}{cccccccc}
\toprule
Sup&Conf&Num$\uparrow$&Hit$\uparrow$&Zero$\downarrow$&Prec$\uparrow$&Recall$\uparrow$&F1$\uparrow$\\
\midrule
10&60\%&1.8&1.3&5.1\%&\textbf{73.0\%}&12.8\%&0.207\\
5&60\%&6.9&4.0&0.5\%&55.3\%&13.4\%&0.207\\
3&60\%&\textbf{25.4}&\textbf{10.3}&0.0\%&40.5\%&\textbf{22.3\%}&\textbf{0.278}\\
3&70\%&15.8&7.2&0.0\%&44.1\%&16.2\%&0.228\\
3&80\%&8.4&4.0&0.0\%&44.7\%&10.9\%&0.169\\
\bottomrule
\end{tabular}
\setlength{\belowcaptionskip}{10pt}
\label{table1}
\end{table}

\subsection{Hyperparameter Threshold Tuning}
The thresholds for support and confidence during rule mining have a strong effect on the credibility of the rules. So we test the performance of different threshold combinations to find the best setting. The value of support is small thus we represent supp($X$) by the number of transactions containing item $X$ for convenience since the denominator of the supp($X$) is a constant.
%, the total number of the transactions. 
As show in Table \ref{table1}, we evaluated the rules in terms of average number of inferred concepts, average hit number, proportion of zero inference photo stream, precision, recall and Macro-F1. The average hit number means the average number of semantic concepts in a photo stream. The values are obtained by testing on 1,000 sampled photo streams. The recall is relatively low because all semantic words in annotations are seen as labels. We find that higher values of two thresholds will lead to more representative and discriminative rules but less inferred concepts, as shown in Table \ref{table1}. This will result in higher precision and lower recall. The value of the thresholds should be set to keep a balance between the quantity and quality of inferred concepts. We choose the values with higher F1 under the premise that there is at least one inferred concept for each photo stream.  Finally $\text{supp}_{min}$ and $\text{conf}_{min}$ are set to 3(equivalent to 0.015$\%$ in proportion form) and 60$\%$. There are 246,197 rules and 1,099 concepts categories in total. The number of inferred concepts for each story is 25.4. We also give out the abstract concepts inferred by mined rules which shares the same patterns as shown in Figure \ref{figure4}.

\subsection{Visualization of Rules}
In order to show the interpretation of the cross-modal rules, we visualize the activated filters of a single rule, as shown in Figure \ref{figure5}. The upper row is the visualization of feature maps of the corresponding filters. The lower row is generated by applying the interpolation of the feature map as a mask to the origin picture. We can find that filter 458, 646, 1484 are detecting round shape, lantern eyes, and pumpkin pedicle respectively; filter 853, 1287 are detecting background. By analysing activated filters combination in the rule, we are able to better understand how the model infers these concepts.

\begin{table*}[t]
\centering
\caption{Automatics metric evaluation. Approaches with * are evaluated in photo stream level, while the others are evaluated in album level. Thus scores of approaches with * are provided only for reference.}
\begin{tabular}{llcccccc}
\toprule
Evaluation&Method&BLUE1-1&BLUE-4&METEOR-v2&ROUGE&CIDEr\\
\midrule
photo stream&Grounded$^*$\cite{huang2016visual}&-&-&31.4&-&-\\
photo stream&HSRL-16topics$^*$\cite{Huang18}&-&11.64&35.0&30.6&-\\
photo stream&HSRL-64topics$^*$\cite{Huang18}&-&13.41&35.2&\textbf{30.8}&-\\
album&h-attn-rank\cite{yu2017hierarchically}&-&-&34.1&29.5&7.5\\
album&CIDEr-RL\cite{wang2018no}&61.9&13.8&34.9&29.7&8.1\\
album&GAN\cite{wang2018no}&62.8&14.0&35.0&29.5&9.0\\
album&AREL\cite{wang2018no}&63.8&14.1&35.0&29.5&\textbf{9.4}\\
\midrule
album&Ours$($baseline$)$&62.5&13.8&35.0&29.6&8.7\\
album&Ours$($VSCMR$)$&\textbf{63.8}&\textbf{14.3}&\textbf{35.5}&30.2&9.0\\
\bottomrule
\end{tabular}
\setlength{\belowcaptionskip}{10pt}
\label{table2}
\end{table*}

\subsection{Automatic Metrics Evaluation}
%In this section, we evaluate our model on automatic metric.
\textbf{Metrics} We first evaluate our approach on automatic metrics. For fair comparison, we utilize the open source evaluation code used in \cite{yu2017hierarchically} \cite{wang2018no}. Scores on BLUE\cite{Papineni02}, METEOR\cite{SATANJEEV05}, ROUGE-L\cite{Lin04}, CIDEr\cite{vedantam2015cider} metrics are reported. We set METEOR score as the main indicator since \cite{huang2016visual} concludes that METEOR correlates best with human judgement according to correlation coefficients.

\textbf{Baseline} We build our baseline by moving out semantic context vector module. The baseline is compared to verify the effectiveness of the proposed cross-modal rules. We provide performances of previous methods for reference. \cite{huang2016visual} uses a separate caption model to generate word candidates to reduce the search space. \cite{yu2017hierarchically} propose an album encoder and a photo selector to encode the album photos and select the most representative photos. \cite{wang2018no} use different metric scores as reward for RL to train the story model. They analyse the limitation of automatic metrics and further design an adversarial reward leaning~(AREL) framework. They also build a model trained by GAN for comparison. \cite{Huang18} use a hierarchically structured reinforcement learning~(HSRL) to improve performance. They design a manager decoder to generate a topic distribution for the lower level worker decoder. 

\textbf{Quantitative analysis} As shown in Table \ref{table2}, our visual storytelling with cross-modal rules~(VSCMR) model achieves the state-of-the-art performance on most of the metrics. We observe perfomance gain on all of the metrics compared to our baselines: 2.1\%, 3.6\%, 1.4\%, 2\%, 3.4\% for BLUE-1, BLUE-4, METEOR, ROUGE and CIDEr respectively. This indicates the effectiveness of the inferred semantic concepts. We optimize our model mainly upon METEOR score since it correlates best with the human judgement according to correlation coefficients\cite{huang2016visual}. We also observe a CIDEr score gap among different papers and investigate into it. The evaluation code of \cite{yu2017hierarchically} evaluates the model on the album level. It takes all stories about the same album as references though their photo streams are not always the same. Meanwhile, \cite{Huang18} evaluates the model using the code released by COCO evaluation server. It only takes stories of the same photo stream as references strictly, which will lead to higher CIDEr score and lower score on other metrics. The difference of the evaluation methods lead to variant score distribution. Thus only methods of \cite{yu2017hierarchically} and \cite{wang2018no} are directly compared with our model for fair comparison. 

\begin{figure}[t] 
\centering 
\includegraphics[height=5.078cm, width=8cm]{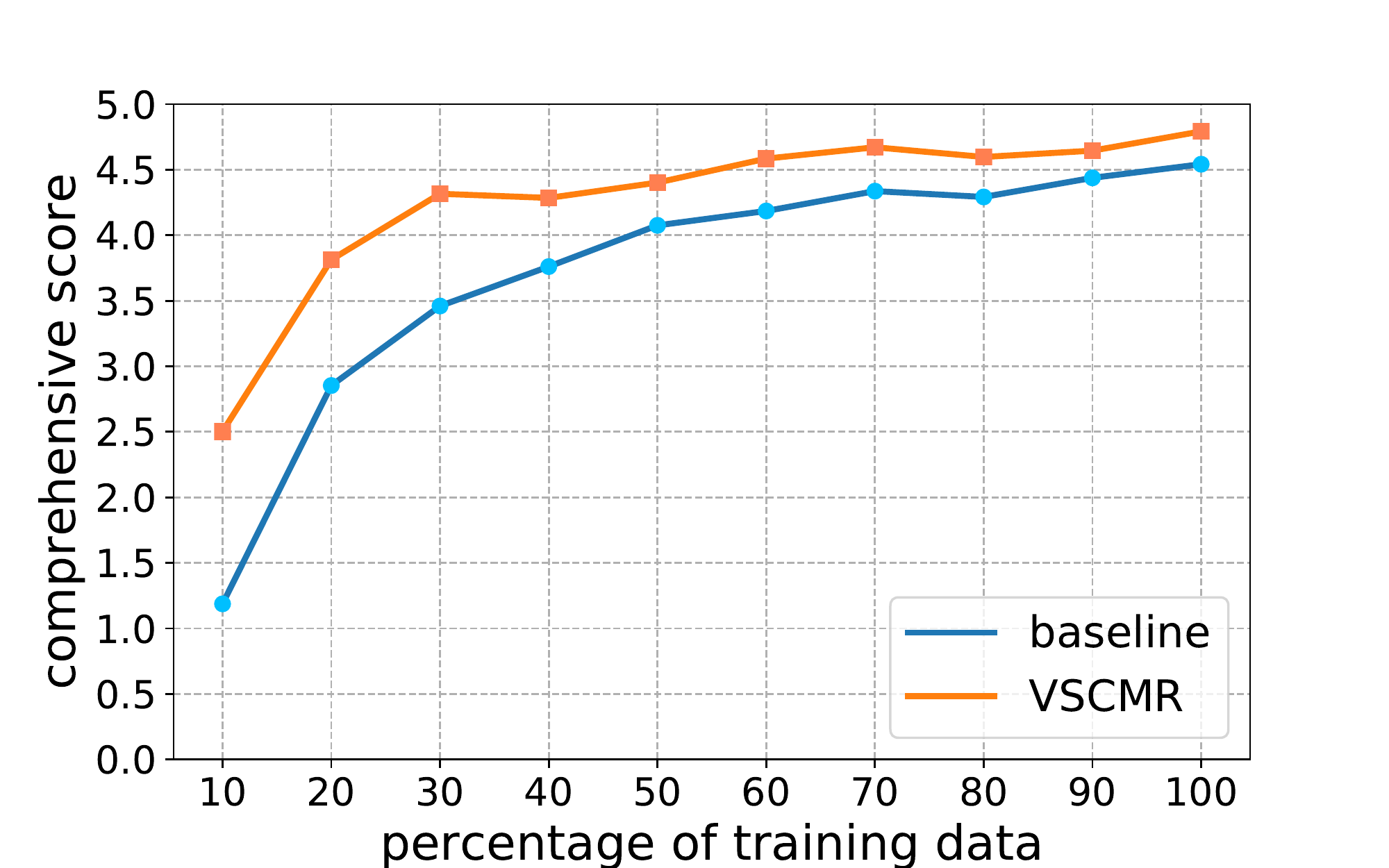}
\caption{Comparison of comprehensive scores on different size of training data. The comprehensive score is the sum of normalized scores on all metrics. Our cross-modal rules help the model generate better stories with less data.}
\label{figure7}
\end{figure}

\begin{figure*}[t] 
\centering 
\includegraphics[height=5.495cm, width=17.8cm]{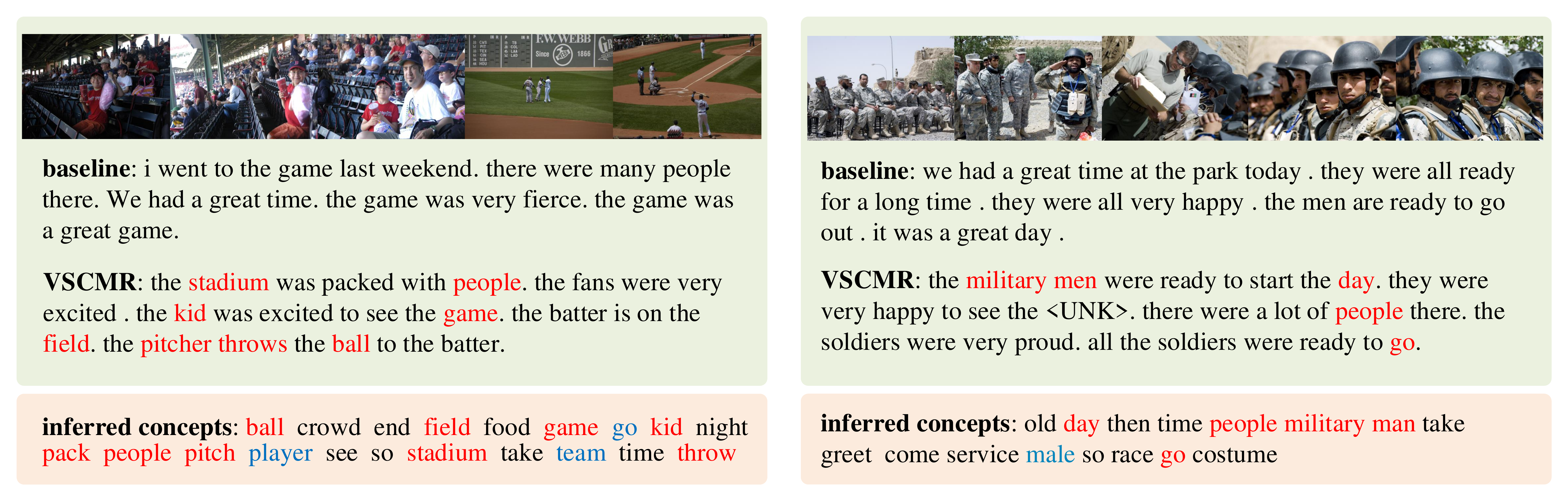} 
\caption{Examples of generated stories. The words in red are inferred concepts which are shown in the model generated story. The words in blue are relative to the photo stream but not shown in the generated story.}
\label{figure8}
\end{figure*}

\subsection{Transaferability study}
In order to demonstrate the transferability of our cross-modal rules, we carried out the ablation study on different size of the training data. We split the training set with a step of 10\%. Both the baseline model and our full model are trained 100 epochs to ensure convergence. To evaluate the overall performance of the model and eliminate metric preference, we calculate a comprehensive score by leveraging all metric scores. For each metric, the score appearing in range of $[l, r]$ is normalized to range of $[0,1]$ by subtracting lower bound $l$ and being divided by the range length $r-l$. The final comprehensive score of the model is the sum of the normalized scores of all metrics. The lower bound is provided by the performance of baseline model trained on 1\% data and the upper bound is the highest score seen in all models. Figure \ref{figure7} shows the performance comparison of the baseline model and VSCMR model. The VSCMR model outperforms the baseline model significantly on small dataset. The performance gain decreases as the size of training data increases since the encoder-decoder model is capable of learning the relationship between vision and language spontaneously with sufficient data. Nevertheless, the VSCMR model has a higher upper bound of the comprehensive score. Our cross-modal rules play a role of external experience to help the model generate better stories with less data.

\begin{table}[t]
\centering
\caption{Pairwise human evaluation. Tie means the worker cannot tell which story is better.}
\begin{tabular}{c|ccc}
\hline
~&\multicolumn{3}{c}{Method}\\
\hline
Choice&baseline&VSCMR&Tie\\
\hline
Relevance&31.4\%&\textbf{45.9\%}&22.7\%\\
Expressiveness&25.6\%&\textbf{45.9\%}&28.5\%\\
Concreteness&19.3\%&\textbf{55.0\%}&25.6\%\\
\hline
\end{tabular}
\setlength{\belowcaptionskip}{10pt}
\label{table3}
\end{table}
\subsection{Human Evaluation}
Considering the limitation of automatic metrics and the complexity of storytelling task, we further perform human evaluation in the form of pairwise comparison. We sample 150 image sequences from the test set. For each image sequence, the human worker is presented with stories generated by the baseline model and the full model and the orders of option within each item is shuffled. The worker is asked to judge which story is better in term of relevance, expressiveness and concreteness. The descriptions of these three evaluation criteria are listed as follows.
\begin{itemize}
    \item \textbf{Relevance} the story accurately describes what is happening in the image sequence and covers the correct topic.
    \item \textbf{Expressiveness} coherence, grammatically and semantically correct, no repetition, expressive language style.
    \item \textbf{Concreteness} the story should narrate concretely what is in the image rather than giving very general descriptions.
\end{itemize}

A "neutral" option is provided if the worker cannot tell which is better. Each sample is assigned to 5 workers to eliminate personal preference.

The result of pairwise comparison is shown in Table \ref{table3}. Our semantic concepts boosted approach significantly outperforms the baseline in terms of concreteness and is better on relevance and expressiveness. The input of the semantic context vector encourages the model to generate the inferred semantic words. This helps the model avoid generating general descriptions and thus improves the concreteness significantly. The concrete descriptions will bring about less repetition which contributes to the relevance and expressiveness.

\subsection{Qualitative Analysis}
To make an intuitive comparison, Figure \ref{figure8} gives the examples of generated stories and inferred semantic concepts. The words in red are inferred concepts which are shown in the model generated story. The words in blue are relative to the photo stream but not shown in the generated story. As shown in Figure \ref{figure8}, the story generated with semantic concepts are more expressive and concrete. The words in the semantic concepts set such as "stadium", "pitcher", "throw", have a greater chance to appear in the description. The words describing events such as "military" will also function as auxiliary information to help the model focus on the related topic, which is crucial for the informative story generation.

\section{CONCLUSION}
We propose a semantic concept inference framework for visual storytelling task by discovering association rules in the cross-modal transactions. The mined cross-modal rules not only help the model to infer semantic concepts including entities, attributes, actions, and events, but also holds the advantages of interpretation, expandability, and transferability as well. Both the automatic metrics and the human evaluation has demonstrated that the inferred semantic concepts help the model generate more informative and grounded stories.

%
% The acknowledgments section is defined using the "acks" environment (and NOT an unnumbered section). This ensures
% the proper identification of the section in the article metadata, and the consistent spelling of the heading.
\begin{acks}
This work has been supported in part by National Key Research and Development Program of China (SQ2018AAA010010), NSFC (No.61751209, U1611461), Hikvision-Zhejiang University Joint Research Center, Zhejiang University-Tongdun Technology Joint Laboratory of Artificial Intelligence, Zhejiang University iFLYTEK Joint Research Center, Chinese Knowledge Center of Engineering Science and Technology (CKCEST), Engineering Research Center of Digital Library, Ministry of Education.
\end{acks}

%
% The next two lines define the bibliography style to be used, and the bibliography file.
\bibliographystyle{ACM-Reference-Format}
\balance
\bibliography{main}

% 
% If your work has an appendix, this is the place to put it.
%\appendix

\end{document}